\documentclass[11pt,twoside]{article}

\begin{document}
\title{\bf Probabilistic exact deletion and probabilistic
no-signalling}
\author{Indranil Chakrabarty$^{1,2}$\thanks{Corresponding
Author:Indranil Chakrabarty ,E-Mail:indranilc@indiainfo.com}, Satyabrata Adhikari $^2$,  B. S. Choudhury $^2$\\
$^1$Heritage Institute of Technology, Kolkata, India\\ $^2$
Department of Mathematics,Bengal Engineering and Science
University, Howrah-711103, \\West Bengal,India} \maketitle
\begin{abstract}
In this work we show that one cannot use non-local resources for
probabilistic signalling even if one can delete a quantum state
with the help of probabilistic quantum deletion machine. Here we
find that probabilistic quantum deletion machine is not going to
help us in identifying two statistical mixture at remote location.
Also we derive the bound on deletion probability from
no-signalling condition.
\end{abstract}
\section{Introduction:}
Entanglement lies in the heart of Quantum mechanics. The key
feature of the entangled states is the non-local correlations
exhibited by them. This nonlocal correlations can not be explained
by any kind of theories like local hidden variable theory [1] etc.
A. Einstein, B.Podolsky and N.Rosen proved that measurement
outcome of one system can instantaneously affect the result of the
other [2]. This observation suggest that one can exploit the non
local nature of the entanglement to send super-luminal signals.
However it was shown that operators at space like separated
distances commute, the average of the observable at distant site
remains the same and does not depend on the operation carried out
by other party [3,4]. Herbert argued that one can amplify the
quantum state on one part of the system and performing
measurements on many copies of that he can get the information
about what the other party has done [5]. Later it was shown that
one cannot clone a single quantum state , popularly known as
'no-cloning' theorem [6].  It was also shown that one cannot clone
two non orthogonal quantum states [7]. Since deterministic exact
cloning is ruled out by linearity of quantum theory, the
possibility of super luminal signalling using non locality of the
entangled states faded away. Various cloning machines which
produce inaccurate copies [8,9,10], cannot be used for super
luminal signalling. It was shown that the fidelity criterion based
on no signalling is consistent with the fidelity criterion of
inaccurate copying [11]. But it was shown by Duan and Guo [12]
that a set of states chosen secretly can be exactly cloned by a
probabilistic cloning machine if and only if the states are
linearly independent. Hardy and Song found out a limit on the
number of states that can be cloned by a probabilistic cloning
machine by using the no signalling criterion [13]. Pati showed
that one cannot send super luminal signals even probabilistically
with the help of a probabilistic quantum cloning machine [14].
Complementary to 'no-cloning' theorem, Pati and Braunstein
introduced a new concept of deletion of an arbitrary quantum state
and shown that an arbitrary quantum states cannot be deleted. This
is due to the linearity property of quantum mechanics.Quantum
deletion [15] is more like reversible 'uncopying' of an unknown
quantum state. The corresponding no-deleting principle does not
prohibit us from constructing the approximate deleting machine
[16].J. Feng et.al. [17] showed that each of two copies of
non-orthogonal and linearly independent quantum states can be
probabilistically deleted by a general unitary-reduction
operation. Like universal quantum cloning machine, D'Qiu [18] also
constructed a universal deletion machine but unfortunately the
machine was found to be non-optimal in the sense of fidelity. A
universal deterministic quantum deletion machine is designed in an
unconventional way that improves the fidelity of deletion from 0.5
and takes it to 0.75 in the limiting sense [16]. Recently we
showed that splitting of quantum information [19] is consistent
with the principle of no-signalling [20]. Pati and Braunstein
showed that if quantum states can be deleted perfectly one can
send signals faster than the speed of light [21]. If a message can
be transmitted instantaneously with a certain no-zero probability
of success less than unit, it is known as probabilistic
super-luminal signalling. It had already been seen that
deterministic as well as probabilistic cloning machine and
deterministic deletion machine does not allow super
luminal signalling between two distant parties.\\\\
The organization of this work is as follows. In section 2, we show
that one cannot send signals faster than light using probabilistic
deletion machine. To show that we have consider a joint system of
two singlet states shared between two distant partners Alice and
Bob. Thereafter, the application of Bob's probabilistic deletion
machine on his qubit leads to superluminal signalling if Alice
prepare her particles in different basis because the resulting
reduced density matrix describing Bob's system will be different
in different basis. However this will not be true. since it is a
known fact that probabilistic deletion machine deletes one copy of
a quantum state from among two copies if and only if the states
are linearly independent, but the states we have considered and
fed into the probabilistic deletion machine are not all linearly
independent so we arrive at a contradiction. In section 3, we
obtain a bound on deletion probability using the no-signalling
condition.
\section{\bf No-signalling using maximally entangled states}
In this section, we show that signals cannot be send faster than
light even with some non-zero probability, using the probabilistic
deletion machine and non-local resources such as maximally
entangled states.\\
Let us consider two maximally entangled states $|\chi\rangle_{12}$
and $|\chi\rangle_{34}$, each prepared in a
singlet state shared by two distant partners, say Alice and Bob.\\
The singlet state $|\chi\rangle_{12}$ can be written in terms of
the basis
$\{|\psi_{i}\rangle,\overline{|\psi_{i}\rangle}\}$(i=1,2) onto
which Alice might do a measurement and therefore the state
$|\chi\rangle_{12}$ is given by
\begin{eqnarray}
|\chi\rangle_{12}=\frac{1}{\sqrt{2}}(|\psi_{i}\rangle\overline{|\psi_{i}\rangle}-
\overline{|\psi_{i}\rangle}|\psi_{i}\rangle)
\end{eqnarray}
Similarly, the entangled state $|\chi\rangle_{34}$ looks exactly
the same as $|\chi\rangle_{12}$. The particles 1 and 3 possessed
by Alice and particles 2 and 4 belongs to Bob.\\
The state representing the combined system
$|\chi\rangle_{12}\otimes |\chi\rangle_{34}$ for arbitrary qubit
basis $\{|\psi_{i}\rangle,\overline{|\psi_{i}\rangle}\}$(i=1,2) is
given by
\begin{eqnarray}
|\chi\rangle_{12}\otimes|\chi\rangle_{34}&=&
\frac{1}{2}(|\psi_{i}\rangle_{1}|\psi_{i}\rangle_{3}|\overline{\psi_{i}}\rangle_{2}
|\overline{\psi_{i}}\rangle_{4}+|\overline{\psi_{i}}\rangle_{1}|\overline{\psi_{i}}\rangle_{3}
|\psi_{i}\rangle_{2}|\psi_{i}\rangle_{4}-|\overline{\psi_{i}}\rangle_{1}|\psi_{i}\rangle_{3}
|\psi_{i}\rangle_{2}|\overline{\psi_{i}}\rangle_{4}-{}\nonumber\\&&
|\psi_{i}\rangle_{1}|\overline{\psi_{i}}\rangle_{3}|\overline{\psi_{i}}\rangle_{2}|\psi_{i}\rangle_{4})
\end{eqnarray}
A probabilistic quantum deleting machine (PQDM) which will delete
one copy among two copies of a quantum state chosen from a set
$S=\{ |\psi_i\rangle,~~i=1,2\}$ is defined by
\begin{eqnarray}
U(|\psi_i\rangle|\psi_i\rangle|P_{0}\rangle)=\sqrt{p_{i}}|\psi_i\rangle|\Sigma\rangle|P_0\rangle+\sqrt{1-p_{i}}|\Phi_i\rangle|P_1\rangle
\\
U(|\overline{\psi_i}\rangle|\overline{\psi_i}\rangle|P_{0}\rangle)=\sqrt{p_{\perp
i}}|\overline{\psi_i}\rangle|\Sigma\rangle|P_0\rangle+\sqrt{1-p_{\perp
i}}|\overline{\Phi_i}\rangle|P_1\rangle\\
U(|\psi_i\rangle|\overline{\psi_i}\rangle|P_{0}\rangle)=|X_{1}\rangle|P_{2}\rangle
\\
U(|\psi_i\rangle|\overline{\psi_i}\rangle|P_{0}\rangle)=|X_{2}\rangle|P_{3}\rangle
\end{eqnarray}
where $|\Sigma\rangle$ denote some standard blank state and
$|P_{0}\rangle$ , $|P_{1}\rangle$, $|P_{2}\rangle$ and
$|P_{3}\rangle$ are probe states to tell us whether deletion has
been successful or not. Here $|P_{1}\rangle$, $|P_{2}\rangle$,
$|P_{3}\rangle$are not necessarily orthogonal, but each of them
are orthogonal to $|P_{0}\rangle$. Here $|\Phi_i\rangle,
|\overline{\Phi_i}\rangle$ are the composite system of the input
state and ancilla whereas $|X_{1}\rangle$ and $|X_{2}\rangle$ are
in general pure entangled states of the non-identical inputs and
the ancilla. We also note that $\{|\phi_i\rangle,
|\overline{\phi_i}\rangle\}$ are not necessarily orthonormal. Here
$p_i$ is the probability of successful deletion of quantum states.
The unitary operator U together with a measurement on the probing
device P deletes one of the two copies with probability of success
$p_i$.\\
To investigate the question of the possibility of the superluminal
signalling with non-zero probability, we carry out the action of
probabilistic quantum deletion machine on one part of a composite
system (say, on the state of Bob's particle 4).\\
After operating probabilistic quantum deletion machine (3-6) on
the state of Bob's particle 4, the combined system (2) reduces to
\begin{eqnarray}
&&|\chi\rangle_{12}\otimes|\chi\rangle_{34}|A\rangle_{6}|P_{0}\rangle_{5}\longrightarrow
U_{245}
(|\chi\rangle_{12}|\chi\rangle_{34}|A\rangle_{6}|P_{0}\rangle_{5}
={}\nonumber\\&&\frac{1}{2}[(\sqrt{p_{\perp
i}}|\psi_i\rangle_{1}|\psi_i\rangle_{3}|\overline{\psi_i}\rangle_{2}+\sqrt{p_{i}}|\overline{\psi_i}\rangle_{1}|\overline{\psi_i}\rangle_{3}|\psi_i\rangle_{2})
~|\Sigma\rangle_{4}|P_0\rangle_{5}{}\nonumber\\&&+(\sqrt{1-p_{\perp
i}}~ |\overline{\Phi_i}\rangle_{24}+\sqrt{1-p_{i}}
~|\Phi_i\rangle_{24})~|P_1\rangle_{5}) {}\nonumber\\&&
+|\psi_i\rangle_{1}|\overline{\psi_i}\rangle_{3}|X_1\rangle_{24}|P_2\rangle_{5}
+
|\overline{\psi_i}\rangle_{1}|\psi_i\rangle_{3}|X_2\rangle_{24}|P_3\rangle_{5}]
\end{eqnarray}
where the ancilla $|A\rangle_{6}$ is attached by Bob and the
states $\{|\psi_i\rangle\}$ and $\{|\overline{\psi_i}\rangle\}$ ,
(i=1,2) are linearly independent.\\
Alice may perform a measurement on her particles 1 and 3 in any
one of the two basis
$\{|\psi_{1}\rangle,|\overline{\psi_{1}}\rangle\}$ and
$\{|\psi_{2}\rangle,|\overline{\psi_{2}}\rangle\}$.\\
If Alice finds her particles in the basis
$\{|\psi_{1}\rangle,|\overline{\psi_{1}}\rangle\}$ then the
reduced density matrix of the particles 2 and 4 in Bob's subsystem
is given by
\begin{eqnarray}
\rho_{24}= Tr_{1356}(\rho_{123456})=
\frac{1}{4}[p_{1}|\psi_{1}\rangle\langle\psi_{1}|+p_{\perp1}
|\overline{\psi_{1}}\rangle\langle\overline{\psi_{1}}|]\otimes|\Sigma\rangle\langle\Sigma|
\end{eqnarray}
On the other hand If Alice finds her particles in the basis
$\{|\psi_{2}\rangle,|\overline{\psi_{2}}\rangle\}$ then the
reduced density matrix of the particles 2 and 4 takes the
following form
\begin{eqnarray}
\rho_{24}=Tr_{1356}(\rho_{123456})=\frac{1}{4}[p_{2}|\psi_{2}\rangle\langle\psi_{2}|+p_{\perp2}
|\overline{\psi_{2}}\rangle\langle\overline{\psi_{2}}|]\otimes|\Sigma\rangle\langle\Sigma|
\end{eqnarray}
It is clear from equations (4) and (5) that the two statistical
mixtures describing the reduced density matrices of the particles
2 and 4 in Bob's subsystem is different and therefore one can
conclude from here that superluminal signalling is possible i.e.
the two different statistical mixtures of the same particles
allowed Bob to distinguish two preparation stages by Alice. But it
is a known fact that the superluminal signalling is not possible.
The fallacy arises because we treat the four states
$\{|\psi_{1}\rangle,|\overline{\psi_{1}}\rangle,|\psi_{2}\rangle,|\overline{\psi_{2}}\rangle\}$
as linearly independent states, which is not true. Since the
states $\{|\psi_{2}\rangle,|\overline{\psi_{2}}\rangle\}$ can be
written as
\begin{eqnarray}
|\psi_{2}\rangle=\cos\theta|\psi_{1}\rangle+
\sin\theta|\overline{\psi_{1}}\rangle\\
|\overline{\psi_{2}}\rangle=\sin\theta|\psi_{1}\rangle-
\cos\theta|\overline{\psi_{1}}\rangle
\end{eqnarray}
so the set
$\textit{S}=\{|\psi_{1}\rangle,|\overline{\psi_{1}}\rangle,|\psi_{2}\rangle,|\overline{\psi_{2}}\rangle\}$
is linearly dependent and hence Bob's probabilistic quantum
deletion machine cannot delete perfectly one copy from the two
copies of all the four states taken from the set \textit{S} with
non-zero probability. Hence one can easily rule out the
possibility of probabilistic superluminal signalling using
probabilistic quantum deletion machine.
\section{\bf Bounds on deletion probability from no-signalling condition}
A bipartite state can be written as
\begin{eqnarray}
|\psi\rangle_{AB}=\frac{1}{\sqrt{N}}\sum_{k=1}^{N}|u_{k}\rangle|v_{k}\rangle
\end{eqnarray}
where $|u_{i}\rangle$ are orthogonal basis for Alice's Hilbert
space $H_{A}$ and $|v_{i}\rangle$ are non-orthogonal and linearly
independent basis states for Bob's Hilbert space $H_{B}$.\\
The combined state of two copies of $|\psi\rangle_{AB}$ is given
by
\begin{eqnarray}
|\psi\rangle_{AB}\otimes|\psi\rangle_{AB}=\frac{1}{N}[\sum_{k=1}^{N}|u_{k}
u_{k}\rangle|v_{k}v_{k}\rangle+\sum_{k\neq l}|u_{k}
u_{l}\rangle|v_{k}v_{l}\rangle]
\end{eqnarray}
After the application of probabilistic quantum deletion machine on
the two copies of non-orthogonal basis states $|v_{i}\rangle$, the
combined state reduces to
\begin{eqnarray}
|\xi\rangle_{ABCD}&=&\frac{1}{N}[\sum_{k=1}^{N}\{|u_{k}
u_{k}\rangle(\sqrt{p_{k}}|v_{k}\rangle|\Sigma\rangle|P_{0}\rangle+
\sqrt{1-p_{k}}|\phi_{k}\rangle|P_{1}\rangle)\}+\sum_{k\neq
l,k<l}|u_{k}
u_{l}\rangle|X_{1}\rangle|P_{2}\rangle{}\nonumber\\&&+\sum_{k\neq
l,k>l}|u_{k} u_{l}\rangle|X_{2}\rangle|P_{3}\rangle]
\end{eqnarray}
Now to calculate the success rate of deleting a copy perfectly
using probabilistic quantum deletion machine, we define a basis in
Bob's Hilbert space as
\begin{eqnarray}
|\zeta_{i}\rangle=\langle u_{i}u_{i}|\xi\rangle_{ABCD}=
\frac{1}{N}[\sqrt{p_{i}}|v_{i}\rangle|\Sigma\rangle|P_{0}\rangle+
\sqrt{1-p_{i}}|\phi_{i}\rangle|P_{1}\rangle]
\end{eqnarray}
Taking the inner product of the two distinct basis, we get
\begin{eqnarray}
\langle\zeta_{i}|\zeta_{j}\rangle=\frac{1}{N^{2}}[\sqrt{p_{i}p_{j}}\langle
v_{i}|v_{j}\rangle+\sqrt{(1-p_{i})(1-p_{j})}\langle\phi_{i}|\phi_{j}\rangle]
\end{eqnarray}
Taking modulus both sides and using the inequality
$|x+y|\leq|x|+|y|$, we get
\begin{eqnarray}
N^{2}|\langle\zeta_{i}|\zeta_{j}\rangle|\leq
\sqrt{p_{i}p_{j}}|\langle
v_{i}|v_{j}\rangle|+\sqrt{(1-p_{i})(1-p_{j})}|\langle\phi_{i}|\phi_{j}\rangle|
\end{eqnarray}
Further using the inequalities
$\sqrt{p_{i}p_{j}}\leq\frac{p_{i}+p_{j}}{2}$,
$\sqrt{(1-p_{i})(1-p_{j})}\leq(1-\frac{p_{i}+p_{j}}{2})$ and
$|\langle\phi_{i}|\phi_{j}\rangle|\leq1$, we have
\begin{eqnarray}
\frac{p_{i}+p_{j}}{2}\leq\frac{1-N^{2}|\langle\zeta_{i}|\zeta_{j}\rangle|}{1-|\langle
v_{i}|v_{j}\rangle|}
\end{eqnarray}
It is clear from equation (18) that the average success
probability of deletion of a qubit depends on the number of the
linearly independent states that can be deleted and also on the
inner product of the non-orthogonal states belonging to Bob's
Hilbert space.
\section{Conclusion}In summary we can say that in this work we
have shown that even though probabilistic quantum deleting machine
can delete one of the two copies at the input port with a non-zero
probability , but  still it is not possible to send super-luminal
signal with certain non-zero probability. Thus  we see that the
quantum theory is in total agreement with the special theory of
relativity.
\section{Acknowledgement}
I.C thank Prof C.G.Chakraborti, for providing encouragement and
inspiration in completing this work. S.A gratefully acknowledge
the partial support by CSIR under the project
F.No.8/3(38)/2003-EMR-1, New Delhi.
\section{Reference}
$[1]$ M. Redhead, \textit{ Incompleteness Nonlocality And Realism
}, (Oxford University Press, Oxford, 1987)\\
$[2]$ A. Einstein, B.Podolosky and N. Rosen, Phys. Rev.
\textbf{47}, 777 (1935).\\
$[3]$ P.Everhard, Nuovo Cim. \textbf{46} B, 392 (1978).\\
$[4]$ G.C.Girihadi.et.al, Lett Nuovo Cim \textbf{27}, 293
(1980).\\
$[5]$ N.Herbert, Found. Phys \textbf{12}, 1171 (1982)\\
$[6]$ W.K.Wooters and W.H.Zurek, Nature \textbf{299}, 802 (1982)\\
$[7]$ H.Yuen, Phys. Lett. A \textbf{113}, 405 (1986)\\
$[8]$ V.Bu$\breve{z}$ek, M.Hillery, Phys. Rev.A \textbf{54}, 1844
(1996)\\
$[9]$ D.Bru$\beta$ et.al, Phys. Rev.A \textbf{57}, 2368
(1998)\\
$[10]$ S.Adhikari, A.K.Pati, I.Chakrabarty, B.S.Choudhury,
\textit{Hybrid
Cloning Machine} (under preparation).\\
$[11]$ N.Gisin, Phys.Lett. A \textbf{242},1 (1998)\\
$[12]$ L.M.Duan and G.C.Guo Phys .Rev. Lett. \textbf{80}, 4999
(1998)\\
$[13]$ L.Hardy and D.D.Song, Phys.Lett. A259 (1999) 331
\\
$[14]$ A.K.Pati, Phys. Lett. A 270 (2000) 103. \\
$[15]$ A.K.Pati and S.L.Braunstein,Nature \textbf{404},164(2000)\\
$[16]$ S. Adhikari, Phys. Rev. A 72, 052321 (2005\\
$[17]$ J.Feng.et.al. Phys. Rev.A \textbf{65}, 052311 (2002)\\
$[18]$ D.Qiu, Phys.Lett.A  \textbf{301},112 (2002)\\
$[19]$ Duanlu Zhou, Bei Zeng, and L. You, \textit{Quantum
information
cannot be split into complementary parts} ,quant-ph/0503168\\
$[20]$ I.Chakrabarty, S.Adhikari, Prashant, B.S.Choudhury ,
\textit{Inseparability of Quantum Parameters} (communicated)\\
$[21]$ A.K.Pati, S.L.Braunstein, Phys. Lett A \textbf{315},
208(2003)
\end{document}